\documentclass[12pt,a4paper]{article}
\usepackage[T1]{fontenc}
\usepackage[sc,osf]{mathpazo}
\usepackage{a4wide}  
\usepackage{amsfonts}
\usepackage{amssymb}
\usepackage{amsmath}
\usepackage{bbm}
\usepackage{wasysym}
\usepackage{booktabs} 
\usepackage{ifpdf}
\ifpdf
\usepackage[pdftex,unicode,implicit]{hyperref}
\hypersetup{%
  pdftitle    = {A non-Abelian Black Ring}
  pdfkeywords = {gravity, supergravity, black hole, black ring, monopole, , 
multicenter, Yang-Mills,  Einstein-Yang-Mills, Higgs, non-Abelian, singular 
monopole, 't~Hooft-Polyakov monopole, Bogomol'nyi, instanton,
Belavin-Polyakov-Schwarz-Tyupkin instanton, coloured, Protogenov, BMPV},
  pdfauthor   = {Tomas Ortin and Pedro F. Ramirez},
  plainpages  = true,
  colorlinks  = true,
  citecolor   = blue,
  urlcolor    = red,
  linkcolor   = black
}
\newcommand{\hepth}[1]{{\tt
\href{http://www.arXiv.org/abs/hep-th/#1}{hep-th/#1}}}

\newcommand{\arxiv}[1]{{\tt arXiv:\href{http://www.arXiv.org/abs/#1}{#1}}}
\newcommand{\doi}[1]{{\tt DOI:\href{http://dx.doi.org/#1}{#1}}}
\else
  \usepackage[dvips]{graphicx}
  \usepackage[unicode,implicit]{hyperref}
  \newcommand{\hepth}[1]{{\tt hep-th/#1}}

  \newcommand{\arxiv}[1]{{\tt arXiv:#1}}
  \newcommand{\doi}[1]{{\tt DOI:#1}}
\fi
\makeatletter
\@addtoreset{equation}{section}
\makeatother

\begin{document}

\begin{flushright}
\small
IFT-UAM/CSIC-16-037\\
April 29\textsuperscript{th}, 2016\\
\normalsize
\end{flushright}

\vspace{2.5cm}

\begin{center}

{\Large {\bf A non-Abelian Black Ring}}
 
\vspace{1.5cm}

\renewcommand{\thefootnote}{\alph{footnote}}
{\sl\large Tom\'{a}s Ort\'{\i}n $^{\text{\quarternote}}$}\footnotetext{E-mail: {\tt Tomas.Ortin [at] csic.es ; p.f.ramirez [at] csic.es}}
{\sl\large  and Pedro F.~Ram\'{\i}rez $^{\text{\quarternote},}$$^{\text{\eighthnote}}$}

\setcounter{footnote}{0}
\renewcommand{\thefootnote}{\arabic{footnote}}

\vspace{1.5cm}

{\it $^{\text{\quarternote}}$Instituto de F\'{\i}sica Te\'orica UAM/CSIC\\
C/ Nicol\'as Cabrera, 13--15,  C.U.~Cantoblanco, E-28049 Madrid, Spain}\\ 

\vspace{0.3cm}

{\it $^{\text{\eighthnote}}$Institut de Physique Th\'eorique, Universit\'e Paris Saclay, CEA\\ CNRS, F-91191 Gif-sur-Yvette, France} \\

\vspace{2cm}


{\bf Abstract}

\end{center}

\begin{quotation}
  We construct a supersymmetric black ring solution of SU$(2)$
  $\mathcal{N}=1,d=5$ Super-Einstein-Yang-Mills (SEYM) theory by adding a
  \emph{distorted} BPST instanton to an Abelian black ring solution of the
  same theory. The change cannot be observed from spatial infinity: neither
  the mass, nor the angular momenta or the values of the scalars at infinity
  differ from those of the Abelian ring. The entropy is, however, sensitive to
  the presence of the non-Abelian instanton, and it is smaller than that of
  the Abelian ring, in analogy to what happens in the supersymmetric coloured
  black holes recently constructed in the same theory and in
  $\mathcal{N}=2,d=4$ SEYM. By taking the limit in which the two angular
  momenta become equal we derive a non-Abelian generalization of the BMPV
  rotating black-hole solution.
\end{quotation}

\newpage
\pagestyle{plain}


\newpage


\section*{Introduction}


The discovery of black rings by Emparan and Reall in
Ref.~\cite{Emparan:2001wn} showed how two important properties of
4-dimensional asymptotically-flat black holes, uniqueness/no-hair and
spherical topology of the event horizon (which, for the 5-dimensional black
ring, is $S^{2}\times S^{1}$), could be violated in higher
dimensions.\footnote{See, for instance, the reviews
  \cite{Emparan:2006mm,Bena:2007kg,Emparan:2008eg} and references therein.}
For a range of values of the conserved charges (mass, angular momenta) that
may characterize an uncharged black ring, a different black-ring and a
black-hole solutions are also possible. For charged black rings (the first of
which was constructed in Ref.~\cite{Elvang:2003yy}) the non-uniqueness becomes
infinite; for the same conserved electric charges one can construct black
rings with regular horizons with magnetic dipole momenta taking continuous
values in some interval \cite{Emparan:2004wy}.  Despite being innocuous to the
conserved charges, these dipole momenta do contribute to the BH entropy. The
construction of supersymmetric black-ring solutions in minimal
\cite{Elvang:2004rt} or matter-coupled $\mathcal{N}=1,d=5$ supergravity
\cite{Gauntlett:2004wh,Bena:2004de,Elvang:2004ds,Gauntlett:2004qy,Ortin:2004af}
using the general classification of supersymmetric solutions of these theories
started in Ref.~\cite{Gauntlett:2002nw} opened up the possibility of
constructing very general families of black-ring solutions with various kinds
of electric charges and moduli in which these issues could be studied.

The violation of the no-hair conjecture by non-Abelian fields in 4-dimensions
is also a well-known but less stressed fact, perhaps because the first
solutions in which this was observed
\cite{Volkov:1989fi,Bizon:1990sr,kn:KunMuA}, black-hole generalizations of the
``Bartnik-McKinnon particle'' \cite{Bartnik:1988am} with asymptotically
vanishing gauge charges, were purely numerical, which makes more difficult
their study and understanding.\footnote{For a review on hairy and non-Abelian
  black-hole solutions see Ref.~\cite{Volkov:1998cc} or the more recent
  Ref.~\cite{Volkov:2016ehx}.} The first black-holes with non-Abelian hair
(not related to the embedding of an Abelian field into a non-Abelian one
through a singular gauge transformation) given in an analytical form were
found using supersymmetry techniques in the context of $\mathcal{N}=2,d=4$
Super-Einstein-Yang-Mills (SEYM) theory\footnote{This theory is the simplest
  $\mathcal{N}=2$ supersymmetric generalization of the Einstein-Yang-Mills
  theory. This supersymmetrization requires the addition of scalar fields to
  the pure Einstein-Yang-Mills theory in order to complete $\mathcal{N}=2,d=4$
  vector supermultiplets and, often, the addition of full vector
  supermultiplets to fulfill the requirements of Special Geometry. There may
  be more than one way of performing this supersymmetrization. Thus, there are
  more than one $\mathcal{N}=2,d=4$ SEYM theory with gauge group SU$(2)$, for
  instance.  These theories are also known as non-Abelian gauged
  $\mathcal{N}=2,d=4$ supergravity coupled to vector supermultiplets.} in
Refs.~\cite{Huebscher:2007hj} and \cite{Meessen:2008kb} using the general
classification of the timelike supersymmetric solutions of these theories made
in Ref.~\cite{Hubscher:2008yz}. The black-hole solutions constructed in
Ref.~\cite{Meessen:2008kb} include the field of an SU$(2)$ \textit{coloured}
monopole found by Protogenov in \cite{Protogenov:1977tq} which also has
asymptotically vanishing gauge charge. The monopole charge does contribute to
the entropy, though. These black holes, which can be seen as the result of
adding the coloured monopole to a standard black hole with Abelian charges,
modifying the entropy but none of the asymptotic charges, were called
\textit{coloured black holes} and they seem to be ubiquitous
\cite{Meessen:2015nla}.

The results of Ref.~\cite{Hubscher:2008yz} have been used more recently to
construct new single-center and two-center non-Abelian solutions of
$\mathcal{N}=2,d=4$ SEYM models that can be obtained by dimensional reduction
of $\mathcal{N}=1,d=5$ SEYM models\footnote{Again, these are the simplest, but
  not unique $\mathcal{N}=1$ (minimal) supersymmetrizations of the $d=5$
  Einstein-Yang-Mills theory and the supersymmetrization requires the addition
  of, at least, scalars. They also go by the name of non-Abelian-gauged
  $\mathcal{N}=1,d=5$ coupled to vector supermultiplets.} in
Ref.~\cite{Bueno:2014mea}.

One of the main goals of that exercise was to open the possibility for the
construction of the first non-Abelian black-hole solutions in $d=5$ by
oxidation to $d=5$ of those solutions, because the direct construction using
the general classification of timelike supersymmetric solutions of
Refs.~\cite{Bellorin:2007yp,Bellorin:2008we} turns out to be too
complicated. This can only be done for certain models of the lower dimensional
theory. The oxidation itself turned out to be a non-trivial exercise if one
wanted to construct solutions without spatial translation isometries (which
would be black strings instead of black holes), but, as was shown in
Ref.~\cite{Bueno:2015wva}, one can use non-trivial cycles to perform the
reduction and still preserve supersymmetry, basically using Kronheimer's
mechanism \cite{kn:KronheimerMScThesis}. Both kinds of black solutions
(strings and holes) were recently constructed in Ref.~\cite{Meessen:2015enl}.

The $d=5$ non-Abelian black holes constructed there are, again, coloured black
holes, with asymptotically vanishing gauge fields. They can be understood as
the result of adding a BPST instanton to a black hole with Abelian charges,
leaving the mass and electric charges unmodified. Just as in the 4-dimensional
case, the non-Abelian field does contribute to the entropy.  The BPST
instanton field turns out to be related by dimensional redox to the coloured
monopole at the heart of the 4-dimensional coloured black holes.

It is natural to try to see if black-rings also admit the addition non-Abelian
instanton fields and the effect this addition may have on the mass and
entropy. In this paper we are going to construct and study a regular
supersymmetric black-ring solution of $\mathcal{N}=1,d=5$ SEYM with a
\emph{distorted} BPST instanton. We start by reviewing in
Section~\ref{sec-recipe} the recipe that we are going to use to construct
timelike supersymmetric solutions, which was obtained in
Ref.~\cite{Meessen:2015enl}. In Section~\ref{sec:nABR} we will carry out the
construction of the solution after which we will study its regularity and we
will compute its essential properties. In Section~\ref{sec-rotatingbhs} we
will study the limit in which the black ring becomes a non-Abelian rotating
black hole. Our conclusions are in Section~\ref{sec-conclusions}.

\section{The recipe to construct solutions}
\label{sec-recipe}

In Ref.~\cite{Meessen:2015enl} we have found a procedure to construct
systematically timelike supersymmetric solutions admitting an additional
spacelike isometry (with adapted coordinate $z$) of any $\mathcal{N}=1,d=5$
Super-Einstein-Yang-Mills (SEYM) characterized by the tensor $C_{IJK}$ and the
structure constants $f_{IJ}{}^{K}$:\footnote{Our conventions are those of
  Refs.~\cite{Bellorin:2006yr,Bellorin:2007yp} and are based on
  Ref.~\cite{Bergshoeff:2004kh}. The supersymmetric solutions of the most
  general $\mathcal{N}=1,d=5$ supergravity theory including vector
  supermultiplets and hypermultiplets and generic gaugings were characterized
  in Ref.~\cite{Bellorin:2007yp}. The inclusion of tensor supermultiplets was
  considered in Ref.~\cite{Bellorin:2008we}.}

\begin{enumerate}
\item Find a set of $t$- and $z$-independent functions $M,H,\Phi^{I},L_{I}$ and
 1-forms $\omega,A^{I},\chi$ in $\mathbb{E}^{3}$
  satisfying the equations (defined in $\mathbb{E}^{3}$ as well)

\begin{eqnarray}
\label{eq:Mequation}
d\star_{3} d M 
& = &
0\, ,
\\
& & \nonumber \\
\label{eq:Hequation}
\star_{3}dH -d\chi  
& = &
0\, ,
\\
& & \nonumber \\
\label{eq:PhiIequation} 
\star_{3}\breve{\mathfrak{D}} \Phi^{I}
- \breve{F}^{I}
& = &
0\, ,
\\
& & \nonumber \\
\label{eq:LIequation}
\breve{\mathfrak{D}}^{2}L_{I} 
-g^{2} f_{IJ}{}^{L}f_{KL}{}^{M}\Phi^{J}\Phi^{K}L_{M}
& = &
0\, ,  
\\
& & \nonumber \\
\label{eq:omegaequation2}
\star_{3} d \omega
-
\left\{
H dM-MdH
+3\sqrt{2} ( \Phi^{I} \breve{\mathfrak{D}}L_{I}
-L_{I}\breve{\mathfrak{D}} \Phi^{I} )
\right\}
& =&
0\, .
\end{eqnarray}

The first two equations state that $H$ and $M$ are harmonic functions on
$\mathbb{E}^{3}$. Once $H$ is given, the second equation (which is the Abelian
Bogomol'nyi equation on $\mathbb{E}^{3}$ \cite{Bogomolny:1975de}) can be
solved for $\chi$. Eq.~(\ref{eq:PhiIequation}) is the general Bogomol'nyi
equation on $\mathbb{E}^{3}$. In the ungauged (Abelian) directions, it implies
that the $\Phi^{I}$ are harmonic functions on $\mathbb{E}^{3}$ and, once they
are chosen, the corresponding vectors $\breve{A}^{I}$ can be determined. In
the non-Abelian directions, the equation becomes non-linear and one has to
find simultaneously solutions for the functions $\Phi^{I}$ and gauge fields
$\breve{A}^{I}$ through adequate ansatzs or other
methods. Eq.~(\ref{eq:LIequation}) is automatically solved if we choose $L_{I}
\propto \Phi^{I}$ (or zero). Finally, Eq.~(\ref{eq:omegaequation2}) can always
be solved if the other equations are solved (because they solve its
integrability condition), except, perhaps, at the singularities of the
functions where, strictly speaking, the other equations are not solved. In
most cases, the integrability condition can be solved by a choice of
integration constants in the functions $H,M,L_{I},\Phi^{I}$. Then, of course,
one has to integrate explicitly Eq.~(\ref{eq:omegaequation2}) to obtain
$\omega$.

\item Using them, reconstruct the solution's 5-dimensional spacetime fields as
  follows:

\begin{enumerate}

  \item The scalars can be found from this equation for the quotients
    $h_{I}(\phi)/\hat{f}$

\begin{equation}
h_{I}/\hat{f} 
= 
L_{I}+8C_{IJK}\Phi^{J}\Phi^{K}/H\, ,
\end{equation}

\noindent
because there is always a parametrization of the scalar manifold such that

\begin{equation}
\phi^{x} \equiv h_{x}/h_{0}\, .
\end{equation}

With the above equation for the quotients $h_{I}(\phi)/\hat{f}$ one can also
determine the function $\hat{f}$. For the special case of symmetric scalar
manifolds, it is given by\footnote{In this expression, $C^{IJK}\equiv
  C_{IJK}$.}

\begin{equation}
\label{eq:f3symmetric-2}
  \begin{array}{rcl}
\hat{f}^{-3}
& = &
3^{3} C^{IJK}L_{I}L_{J}L_{K}
+3^{4}\cdot 2^{3}  C^{IJK}C_{KLM}L_{I}L_{J}\Phi^{L}\Phi^{M}/H
\\
& & \\
& &
+3\cdot 2^{6}L_{I}\Phi^{I}C_{JKL}\Phi^{J}\Phi^{K}\Phi^{L}/H^{2}
+2^{9}\left(C_{IJK}\Phi^{I}\Phi^{J}\Phi^{K}\right)^{2}/H^{3}\, .
\end{array}
\end{equation}

\item The metric has the form

\begin{equation}
\label{eq:themetric}
ds^{2} 
= 
\hat{f}^{\, 2}(dt+\hat{\omega})^{2}
-\hat{f}^{\, -1}d\hat{s}^{2}\, ,  
\end{equation}

\noindent
where $\hat{f}$ has been determined above, the 1-form $\hat{\omega}$ is given
by\footnote{The unhatted $\omega$ is the one occurring in
  Eq.~(\ref{eq:omegaequation2}).}

\begin{eqnarray}
\label{eq:omegahat}
\hat{\omega} 
& = & 
\omega_{5}(dz+\chi) +\omega\, ,
\\
& & \nonumber \\
\label{eq:omega5}
\omega_{5}
& = &
M+16\sqrt{2} H^{-2} C_{IJK} \Phi^{I} \Phi^{J} \Phi^{K}
+3\sqrt{2} H^{-1} L_I \Phi^{I} \, ,
\end{eqnarray}

\noindent
and where the 4-dimensional Euclidean metric $d\hat{s}^{2}$ is given
by\footnote{With $H$ and $\chi$ related by Eq.~(\ref{eq:Hequation}), this is a
  hyperK\"ahler metric admitting a triholomorphic Killing vector, also known
  as Gibbons-Hawking metric \cite{Gibbons:1979zt,Gibbons:1987sp}. We will also
  denote the compact coordinate $z$ by $\varphi$. It will be assumed to take
  values in $[0,4\pi)$.}

\begin{equation}
\label{eq:GHmetric}
d\hat{s}^{2}
=
H^{-1} (dz +\chi)^{2}
+H dx^{r}dx^{r}\, ,\,\,\,\, r=1,2,3\, .
\end{equation}

\item The vector fields and their corresponding field strengths are given by 

\begin{equation}
\label{eq:vectorfields}
\begin{array}{rcl}
A^{I} 
& = &
-\sqrt{3}h^{I}\hat{f} (dt +\hat{\omega}) +\hat{A}^{I}\, ,    
\\
& & \\
F^{I}
& = &
-\sqrt{3} \hat{\mathfrak{D}}[ h^{I}\hat{f} (dt +\hat{\omega}) ]  +\hat{F}^{I}\, ,
\end{array}
\end{equation}

\noindent
where the vector fields $\hat{A}^{I}$, defined on the 4-dimensional Euclidean
space $d\hat{s}^{2}$, and their field strengths are given by 

\begin{equation}
\label{eq:instantondec}
\begin{array}{rcl}
\hat{A}^{I}
& = &
2\sqrt{6} \left[H^{-1}\Phi^{I} (dz+\chi)-\breve{A}^{I} \right]\, ,
\\
& & \\
\hat{F}^{I}
& = &
2\sqrt{6} H^{-1} 
\left[\breve{\mathfrak{D}} \Phi^{I} \wedge (dz+\chi)
-\star_{3} H \breve{\mathfrak{D}} \Phi^{I} \right] \, ,
\end{array}
\end{equation}

\noindent
where $\hat{\mathfrak{D}}$ (resp.~$\breve{\mathfrak{D}}$) is the exterior
gauge-covariant derivative with respect to the connection $\hat{A}^{I}$
(resp.~$\breve{A}^{I}$).

\end{enumerate}

\end{enumerate}

In Ref.~\cite{Meessen:2015enl} we used this recipe to construct black-hole
solutions with non-Abelian gauge and scalar fields for the SU$(2)$-gauged
ST$[2,5]$ model.\footnote{Actually, this is the name of the model of
  $\mathcal{N}=2,d=4$ supergravity one obtains by dimensional reduction.}
This model has 4 vector multiplets and, hence, 4 scalar fields that
parametrize the symmetric space $\mathrm{SO}(1,3)/\mathrm{SO}(3)$. It is
defined by a tensor $C_{IJK}$ with the following non-vanishing components

\begin{equation}
C_{0xy}= \tfrac{1}{6}\eta_{xy}\, ,
\mbox{where}
\,\,\,\,\,
(\eta_{xy}) = \mathrm{diag}(+-\dotsm -)\, ,
\,\,\,\,\,
\mbox{and}
\,\,\,\,\,
x,y=1,\cdots,4\, .
\end{equation}

The directions to be gauged are the last three, which we will denote by
indices $\alpha,\beta,\ldots=2,3,4$. the ungauged directions will be denoted
by indices $i,j,\ldots=0,1$.

Being a symmetric space, we can use Eq.~(\ref{eq:f3symmetric-2}) to write the
metric function $\hat{f}$ as a function of the building blocks
$H,L_{I},\Phi^{I}$:

\begin{equation}
\label{eq:d5metricfunctionST24}
  \begin{array}{rcl}
\hat{f}^{\, -1}
& = &
H^{-1}
\left\{
\tfrac{1}{4}
\left(
6HL_{0} +8\eta_{xy}\Phi^{x}\Phi^{y}
\right)
\left[
9H^{2}\eta^{xy}L_{x}L_{y} +48 H\Phi^{0}L_{x}\Phi^{x}
\right.
\right.
\\
& & \\
& & 
\left.\left.
+64(\Phi^{0})^{2}\eta_{xy}\Phi^{x}\Phi^{y}
\right]
\right\}^{1/3}\, .
\end{array}
\end{equation}

Now, in order to find solutions of this model, we just need to find building
blocks that satisfy Eqs.~(\ref{eq:Mequation})-(\ref{eq:omegaequation2}). In
the next section we will just do this to find a solution that describes a
black ring.

\section{Non-Abelian Black Rings} 
\label{sec:nABR}

\subsection{Construction of the Solution} 

Inspired by Refs.~\cite{Gauntlett:2004qy,Gauntlett:2004wh}, we choose a point
$\vec{x}_{0}\equiv \left(0,0,-R^{2}/4 \right)$ in $\mathbb{E}^{3}$ and
a harmonic function $N$ with a pole at that point,

\begin{equation}
N \equiv \frac{1}{\vert \vec{x}-\vec{x}_{0} \vert} \equiv \frac{1}{r_{n}} \, ,
\end{equation}

\noindent
in terms of which we can write the non-vanishing building blocks in the
ungauged directions as

\begin{equation}
H=\frac{1}{r} \, , 
\hspace{.5cm}
M=\tfrac{3}{4}\lambda_{i}q^{i}\left(1-\vert \vec{x}_{0} \vert N \right) \, , 
\hspace{.5cm}
\Phi^{i}=-\frac{q^{i}}{4\sqrt{2}} N \, ,
\hspace{.5cm}
L_{i}=\lambda_{i}+\frac{Q_{i}-C_{ijk}q^{j}q^{k}}{4} N \, .
\end{equation}

These functions contain the integration constants $q^{i},Q_{i}$ and
$\lambda_{i}$. The first two can be interpreted as charges. The latter, whose
value will be restricted by requirements such as the normalization of the
metric at infinity, are moduli. Eq.~(\ref{eq:Mequation}) is satisfied
automatically. Eq.~(\ref{eq:Hequation}) is satisfied with

\begin{equation}
\chi
=
\cos{\theta}d\psi\, , 
\end{equation}

\noindent
where $r,\theta \in (0,\pi)$ and $\psi \in [0,2\pi)$ are spherical coordinates
centered at $r=|\vec{x}|=0$ with the definitions and orientation

\begin{equation}
\label{eq:coord0}
\left\{
  \begin{array}{rcl}
x^{1} & = & r\sin{\theta} \sin{\psi} \, , \\
x^{2} & = & r\sin{\theta} \cos{\psi} \, , \\
x^{3} & = & -r\cos{\theta} \, ,\\
\end{array}
\right.
\hspace{1cm}
\epsilon^{123}=\epsilon^{r\theta\psi}=+1 \, .
\end{equation}

\noindent
Eqs.~(\ref{eq:PhiIequation}) are satisfied with 

\begin{equation}
\breve{A}^{i}
=
-\frac{q^{i}}{4\sqrt{2}} \cos{\theta_{n}}d\psi_{n}\, ,  
\end{equation}

\noindent
where $r_{n},\theta_{n} \in (0,\pi)$ and
$\psi_{n} \in [0,2\pi)$ are spherical coordinates centered at
$r_{n}=|\vec{x}_{n}|=0$ with the definitions

\begin{equation}
\label{eq:coordn}
\left\{
  \begin{array}{rcl}
x^{1}_{n} & \equiv & x^{1}- x^{1}_{0} = r_{n}\sin{\theta_{n}} \sin{\psi_{n}} \, , \\
x^{2}_{n} & \equiv & x^{2}- x^{2}_{0} = r_{n}\sin{\theta_{n}} \cos{\psi_{n}} \, , \\
x^{3}_{n} & \equiv & x^{3}- x^{3}_{0} = -r_{n}\cos{\theta_{n}} \, ,\\
\end{array}
\right.
\end{equation}

\noindent
and the same orientation as the spherical coordinates centered at $r=0$.

Eqs.~(\ref{eq:LIequation}) in the Abelian directions are trivially satisfied
because all $f_{ij}\,^{k}=0$ and, finally, the integrability condition of
Eq.~(\ref{eq:omegaequation2}) is identically satisfied for the chosen
integration constants and $\omega$ can be found by integration. We will
compute $\omega$ for the complete solution later.

The above functions are enough to construct an Abelian black ring. Now, we
excite the gauged directions of this solution by adding to it a solution of
the SU$(2)$ Bogomol'nyi equations on $\mathbb{E}^{3}$ (\ref{eq:PhiIequation})

\begin{equation}
\Phi^{\alpha}
=
\frac{1}{gr_{n}\left( 1+\lambda^{2} r_{n} \right)} \delta^{\alpha}_{s+1} \frac{x^{s}_{n}}{r_{n}} \, ,
\hspace{1cm}
\breve{A}^{\alpha} 
= 
\frac{1}{g r_{n} \left(1+\lambda^{2} r_{n} \right)} \epsilon^{\alpha}{}_{rs} 
\frac{x^{s}_{n}}{r_{n}} d x^{r}_{n} \, .
\end{equation}

This solution, originally found by Protogenov in
Ref.~\cite{Protogenov:1977tq}, describes a magnetic colored monopole placed at
$r_{n}=0$. It is singular at $r_{n}=0$ as a field configuration in
$\mathbb{E}^{3}$, but this behaviour can change when we analyze the whole
picture. In fact, we showed in Ref.~\cite{Bueno:2015wva} that the monopole
field gives rise to a BPST instanton in $\mathbb{E}^{4}$ through
(\ref{eq:instantondec}), and we used this result in
Ref.~\cite{Meessen:2015enl} to construct a regular black hole of the same
supergravity theory we consider in this work.

In the present case we obtain a different instanton field configuration from
(\ref{eq:instantondec}), which we call \emph{distorted} BPST, because the pole
of the harmonic function $H$ is placed in a different point ($r=0$) than that
of the coloured monopole ($r_n=0$). This distorted BPST is singular at
$r_n=0$, which might turn the black ring solution ill-defined. Happily this is
not the case. The complete vector field contains the instanton plus an
additional term, see (\ref{eq:vectorfields}), where the latter cancels
precisely this divergence at that critical point

\begin{equation}
\lim_{r_n \rightarrow \infty} \left(-\sqrt{3} h^{I} \hat{f} \omega_{5} 
+ 2\sqrt{6} H^{-1} \Phi^{I} \right) \left(dz+\chi \right) = 0 \, .
\end{equation}

Observe that in the ungauged case the $\Phi^{\alpha}$s would have been
harmonic functions $-q^{\alpha}N/(4\sqrt{2})$ and the combinations
$C_{ijk}q^{j}q^{k}$ should have been replaced by $C_{iJK}q^{J}q^{K}$. Here the
asymptotic behaviour of the non-Abelian gauge field indicates that the
``non-Abelian $q^{\alpha}$s'' do not contribute in the same way the $q^{i}$s
do. However, they have a similar near-horizon behaviour.

The above functions define completely the solution. In what follows we are
going to analyze its metric to show that it describes a regular black ring and
to compute its main properties.

\subsection{Analysis of the Solution} 

In this analysis it is convenient to use two set of coordinates: those
centered at $r=0$, ($r,\theta,\psi$, defined in Eq.~(\ref{eq:coord0}))
supplemented by the time coordinate $t$ and the angular coordinate $\varphi$,
and those centered at $r_{n}=0$ ($r_{n},\theta_{n},\psi_{n}$, defined in
Eq.~(\ref{eq:coordn})) supplemented by the time coordinate $t_{n}$ and the
angular coordinate $\varphi_{n}$. The relations

\begin{equation}
  \begin{array}{rcl}
r_{n}
& = & 
\left(r^{2}+\vert \vec{x}_{0} \vert^{2}
-2\vert \vec{x}_{0} \vert r \cos{\theta} \right)^{1/2} \, , 
\\
& & \\
r
& = &
\left(r_{n}^{2}+\vert \vec{x}_{0} \vert^{2}+2\vert \vec{x}_{0} \vert r_{n}
  \cos{\theta_{n}} \right)^{1/2} \, , 
\\
& & \\
\vert \vec{x}_{0} \vert
& = &
r\cos{\theta}-r_{n}\cos{\theta_{n}}\, ,
\end{array}
\end{equation}

\noindent
will be useful in the computations.

The metric function $\hat{f}$ can be obtained by substituting the functions
$H,L_{I},\Phi^{I}$ in Eq.~(\ref{eq:f3symmetric-2}). At this moment we just
want to impose the standard asymptotic normalization

\begin{equation}
\label{eq:normalization}
  \lim_{r\rightarrow \infty} \hat{f}
  =
  1\, ,
\,\,\,\,\,
\Rightarrow
\,\,\,\,\,
3^{3} C^{ijk} \lambda_{i} \lambda_{j} \lambda_{k}
=
\frac{3^{3}}{2}\lambda_{0}\lambda_{1}^{2}=1 \, .
\end{equation}

Now let us compute the only missing ingredient in the metric
(\ref{eq:themetric}): the 1-form $\hat{\omega}$. Let us consider
Eq.~(\ref{eq:omegaequation2}), which, upon substitution of the chosen
functions $H,M,L_{I},\Phi^{I}$, can be written as

\begin{equation}
  \begin{array}{rcl}
\star_{3} d \omega 
& = &
-\tfrac{3}{4} \lambda_{i} q^{i} 
\bigg\{ 
{\displaystyle
-\frac{1}{r^{2}}\left[1-\frac{\vert \vec{x}_{0} \vert +
    r}{r_{n}}+\frac{r \vert \vec{x}_{0} \vert \left( r+\vert \vec{x}_{0} \vert
    \right)}{r_{n}^{3}} \left(1-\cos{\theta} \right) \right] dr 
}
\\
& & \\
& & 
+ 
{\displaystyle
\left[ \frac{\vert \vec{x}_{0} \vert \sin{\theta}}{r_{n}^{3}} \left(r-\vert
    \vec{x}_{0} \vert \right) \right] d\theta 
}
\bigg\} \, ,
\\
\end{array}
\end{equation}

\noindent
and a solution can be readily found assuming $\omega$ has only one
non-vanishing component, $\omega_{\psi}$:\footnote{The expression coincides
  with that of \cite{Ortin:2015hya} despite we have chosen $\vec{x}_{0}$ to be
  on the negative $x^{3}$ axis. This is because the coordinate $\theta$ has
  also a relative sign with respect to the used in that reference.}

\begin{equation}
\omega
=
-\tfrac{3}{4}\lambda_{i}q^{i}
\left(\cos{\theta}-1\right)
\left[1-\left(r+\frac{R^{2}}{4} \right) \frac{1}{r_{n}} \right] d\psi \, .
\end{equation}

Observe that, since $L_{\alpha}=0$ the non-Abelian terms do not affect
$\omega$. However, they do affect the whole 5-dimensional $\hat{\omega}$
given in Eq.~(\ref{eq:omegahat}) via $\omega_{5}$ in Eq.~(\ref{eq:omega5}):

\begin{eqnarray}
\label{eq:omegahat2}
\hat{\omega}
& = &
\left(F-G\right)d\varphi+\left(F-G\cos{\theta}\right)d\psi \, ,
\\
& & \nonumber \\
\label{eq:F}
F 
& = &
\frac{3\lambda_{i}q^{i}}{4}\left[1-\left(r+\frac{R^{2}}{4} \right) 
\frac{1}{r_{n}} \right]\, ,
\\
& & \nonumber \\
\label{eq:G}
G
& = &
\frac{q^{i} }{16 }\left[3\left(Q_{i}-C_{ijk}q^{j}q^{k}\right)
+2C_{ijk}q^{j}q^{k} \frac{r}{r_{n}} \right] \frac{r}{r_{n}^{2}}
-\frac{2q^{0}}{g^{2}} \frac{r^{2}}{r_{n}^{3} \left(1+\lambda^{2}r_{n}\right)^{2}} \, .
\end{eqnarray}

The last term in $G$ has a non-Abelian origin. In the $r\rightarrow \infty$
limit in which the metric tends to Minkowski's (so we have an asymptotically
flat solution), though, it is subdominant and we do not expect it to
contribute to the the angular momentum of the solution.

So far we have been working in coordinates in which the hyperK\"ahler metric
Eq.~(\ref{eq:GHmetric}) is of the form

\begin{equation}
d\hat{s}^{2}
= 
r \left(d\varphi+\cos{\theta} d\psi \right)^{2}
+\frac{1}{r}
\left[
dr^{2}+r^{2}\left(d\theta^{2}+\sin{\theta}^{2}d\psi^{2}\right) 
\right] \, ,
\end{equation}

\noindent
but, in order to compute mass and angular momentum, it is convenient to use a
different coordinate system (also centered at $\vec{x}=0$)
$t,\Theta,\phi_{1},\phi_{2}$, related to the former by

\begin{equation}
r=\frac{\rho^{2}}{4} \, , 
\hspace{1cm} 
\theta=2\Theta \, , \hspace{1cm} 
\psi=\phi_{1}-\phi_2 \, , \hspace{1cm} 
\varphi=\phi_{1}+\phi_2 \, ,
\end{equation}

\noindent
in which the complete 5-dimensional metric is of the form

\begin{equation}
ds^{2}
=
\hat{f}^{2}  \left(dt+\hat{\omega} \right)^{2}
-\hat{f}^{-1} 
\left[ 
d\rho^{2}
+\rho^{2}\left(d\Theta^{2}+\cos^{2}\Theta d\phi_{1}^{2}+\sin^{2}\Theta
  d\phi_{2}^{2} \right) 
\right] \, ,
\end{equation}

\noindent
with

\begin{equation}
\hat{\omega}
=
\left(2F-2G\cos^{2}\Theta\right)d\phi_{1}
-2G\sin^{2}\Theta d\phi_{2}\, .  
\end{equation}

\noindent
The independent components of the angular momentum are now obtained from the
metric behaviour in the $\rho\rightarrow \infty $ limit\footnote{We use units
  in which $G_{N}=\sqrt{3}\pi/4$.}

\begin{eqnarray}
J_{\phi_{1}}
& = &
\lim_{\rho\rightarrow\infty}\frac{\pi \vert g_{t\phi_{1}}\vert \rho^{2}}{4
  G_{N} \cos^{2}{\Theta}}
 =\tfrac{1}{2\sqrt{3}}q^{i}\left(3Q_{i}-C_{ijk}q^{j}q^{k}\right)\, ,
\\
& & \nonumber \\
J_{\phi_{2}}
& = &
\lim_{\rho\rightarrow\infty}\frac{\pi \vert g_{t\phi_{2}} \vert \rho^{2}}{4
  G_{N} \sin^{2}{\Theta}}
=
\tfrac{1}{2\sqrt{3}}q^{i}\left(3Q_{i}
-C_{ijk}q^{j}q^{k}+6\lambda_{i}R^{2}\right) \, ,
\end{eqnarray} 

\noindent
and, from the absence of contribution proportional to $g$, we see that they
coincide with those of the Abelian black ring, as we expected. 

Observe that these formulae allow us to identify

\begin{equation}
\label{eq:modu}
 q^{i}\lambda_{i}R^{2} = \tfrac{1}{\sqrt{3}} (J_{\phi_{2}}-J_{\phi_{1}})\, . 
\end{equation}

Before we move to study the possible presence of an event horizon, let us
point out that the solution does not contain any Dirac-Misner
strings.\footnote{They could have been removed but only at the price of
  introducing closed timelike curves \cite{Misner:1963fr}.} Indeed, the
$g_{t\phi_{1}}$ (resp.~$g_{t\phi_{2}}$) metric component vanishes when the
coordinate $\phi_{1}$ (resp.~$\phi_{2}$) is not well defined, which happens at
$\Theta=\pi/2$ ($\Theta=0$).

The solution may have an event horizon at $\vec{x}=\vec{x}_{0}$, where the
norm of the timelike Killing vector of the metric vanishes. In order to study the near
horizon limit we need to use a different coordinate system because several
components of the metric blow up there in the coordinates we
have been using so far. Recall the expression for the metric in the
original frame centered at $\vec{x}=0$

\begin{equation}
  \begin{array}{rcl}
ds^{2}
& = &
\hat{f}^{2} \left( dt+\omega \right)^{2}+2\hat{f}^2 \omega_{5}
(dt+\omega)(d\varphi+\cos{\theta}
d\psi)
\\
& & \\
& & 
-\hat{f}^{2}\left(\hat{f}^{-3}r-\omega_{5}^{2}\right) 
\left(d\varphi+\cos{\theta}d\psi \right)^{2}
-\hat{f}^{-1}r^{-1}dx^{r} dx^{r} \, .
\end{array}
\end{equation}

\noindent
We first go to the auxiliary frame centered at the horizon with spherical
coordinates and take the $r_{n}\rightarrow 0$ limit. The functions that appear
in the metric behave in this limit as follows

\begin{eqnarray}
\hat{f} 
& = &
\frac{16}{R^{2} v^{2}} r^{2}_{n} + \mathcal{O}(r^{3}_{n}) \, , 
\\
& & \nonumber \\
\omega_{\psi_{n}}  
& = &
-\frac{3}{R^{2}}\lambda_{i} q^{i} \sin^{2}{\theta_{n}} r_{n}
+\mathcal{O}(r_{n}^{2}) \, 
\\
& & \nonumber \\
\hat{f}^{-1}r^{-1}  
& = &
\frac{v^{2}}{4} r_{n}^{-2}+k_{1} r_{n}^{-1}+\mathcal{O}(r_{n}) \, , 
\\
& & \nonumber \\
\hat{f}^{2} \omega_{5} 
& = &
-\frac{2}{v} r_{n} + k_{2} r_{n}^{2}+\mathcal{O}(r_{n}^{3}) \, , 
\\
& & \nonumber \\
\hat{f}^{2} (\hat{f}^{-3} r -\omega_{5}^{2} ) 
& = &
\frac{l^{2}}{4}+k_{3} r_{n}+\mathcal{O}(r_{n}^{2}) \, ,
\end{eqnarray}

\noindent
where we have defined the constants

\begin{eqnarray}
\label{eq:v}
v 
& = &
\left( C_{ijk}q^{i}q^{j}q^{k}-16\frac{q^{0}}{g^{2}} \right)^{1/3} \, , \\ 
& & \nonumber \\
\label{eq:l}
l 
& = &
\frac{1}{2v^{2}} \left[ 9 \cdot 6^{2} C^{ijk} C_{klm} 
\left( Q_{i}-C_{ihn} q^{h} q^{n} \right) 
\left( Q_{j}-C_{jpq} q^{p} q^{q} \right) q^{l} q^{m}
 \right. 
\nonumber \\
& & \nonumber \\
& & 
\left. 
- 9 \left( q^{i} Q_{i}-C_{ijk}q^{i} q^{j} q^{j} \right)^{2} 
- 12 q^{i} \lambda_{i} R^{2} v^{3}
- 9 \left( Q_{1}- \frac{q^{0} q^{1}}{3}  \right)^{2} 
\left( \frac{32}{g^{2}} \right) \right]^{1/2} \, .
\end{eqnarray}

These expression for the constants $v$ and $l$ resemble those of the Abelian
case \cite{Gauntlett:2004qy}, with an additional non-Abelian term. The precise
form of the constants $k_{1}$, $k_{2}$ and $k_{3}$ in terms of the charges are
messy. They do not occur in the calculation of any physical quantity, but they
play a role in the near horizon analysis,\footnote{We give
  their form here for the sake of completeness,
\begin{eqnarray}
k_{1} 
& = & 
\frac{16 \lambda^{2} R^{2} \frac{q^{0}}{g^{2}} +3\left( q^{i} Q_{i}-C_{ijk}
    q^{i} q^{j} q^{k} \right)}{3R^{2} v} \, , \\
k_{2} 
& = & 
\frac{4 k_{1}}{v} \, , \\ 
k_{3} 
& = & 
\frac{1}{2 g^{2} R^{2} v^4} 
\bigg\{ 
\frac{3 R^{2} k_{1}}{v^{3}} \left[ (q^{0}
  q^{1}/3)^{2} \left(96-3 g^{2} (q^{1})^{2}\right)
\right.
\nonumber \\
& &
\hspace{1.5cm}
+6  (q^{0} q^{1}/3) \left(-32 Q_{1}+g^{2} q^{1} 
(-2 (q^{1})^{2}/6 q^{0}+2 q^{0}  Q_{0}+q^{1}Q_{1})\right) 
\nonumber \\
& &
\left.  \hspace{1.5cm}
+3 \left(4 (q^{1})^{2}/6 g^{2} q^{0} q^{1} Q_{1}+32 Q_{1}^{2}
+g^{2} \left(-q^{1}
    Q_{1} (4 q^{0} Q_{0}+q^{1}Q_{1})+ (C_{ijk}q^{i} q^{j} q^{k}-q^{i}
    Q_{i})^{2}\right)\right)  \right. 
\nonumber \\ 
& &
\left.  \hspace{1.5cm}
+4 \lambda_{i} q^{i} \left(-16 q^{0}+ g^{2}  C_{ijk} q^{i} q^{j} q^{k}\right)
R^{2}\right] -3 \left[9 g^{2} ((q^{1})^{2}/6-Q_{0}) (q^{0}q^{1}/3-Q_{1})^{2} 
\right.  
\nonumber \\ 
& &
\left. \hspace{1.5cm} 
+\left(-24 (q^{0}q^{1}/3)^{2} \lambda^{2}
+6 (q^{1})^{2}/6 g^{2} \lambda_{1} q^{0}
  q^{1}-6 g^{2} \lambda_{1} q^{0} Q_{0} q^{1} +96 \lambda_{1} Q_{1}-6 g^{2}
  \lambda_{0} q^{0} q^{1}Q_{1}
\right. \right. 
\nonumber \\ 
& &
\left. \left.  \hspace{1.5cm}
-3 g^{2}\lambda_{1} (q^{1})^{2} Q_{1}
-24 \lambda^{2} Q_{1}^{2}+3 q^{0}q^{1}/3 \left(-32 \lambda_{1}+2 g^{2}
  \lambda_{0} q^{0} q^{1}+g^{2} \lambda_{1} (q^{1})^{2}+16 \lambda^{2}
  Q_{1}\right)  
\right.  \right. 
\nonumber \\ 
& &
\left.  \left.  \hspace{1.5cm}
+32 \lambda_{i} q^{i} q^{0}
-8 C_{ijk} q^{i} q^{j} q^{k} g^{2} \lambda_{i} q^{i}
+6 g^{2}\lambda_{i} q^{i} Q_{j} q^{j}\right) R^{2}
+16 \lambda^{2}  \lambda_{i} q^{i} q^{0} R^4\right] 
   \bigg\}
\end{eqnarray}
} since they are responsible for the disappearance of
$\mathcal{O}(r_{n}^{-1})$ in the metric after we perform the following
coordinate transformation,

\begin{equation}
\label{eq:coordhorizon}
dt_{n}
=
d\tau_{n}+\left( \frac{b_{2}}{r_{n}^{2}}
+\frac{b_{1}}{r_{n}} \right) dr_{n}\, , 
\hspace{1cm}
d\varphi_{n}=-d\psi_{n}+2d\xi_{n}+\frac{c_{1}}{r_{n}}dr_{n} \, ,
\end{equation}

\noindent
where the constants $b_{1}$, $b_{2}$ and $c_{1}$ can be chosen such that all
divergences in the metric in the $r_{n}\rightarrow 0$ limit disappear:

\begin{equation}
c_{1}=\mp \frac{v}{l} \, , 
\hspace{1cm} 
b_{2}=\pm \frac{lv^{2}}{8} \, , 
\hspace{1cm}
b_{1}=\pm \frac{4l^{2} k_{1}+ l^{2} v^{3} k_{2} + 4 v^{2} k_{3}}{16l} \, .
\end{equation}

With this choice we find in the $r_{n}\rightarrow 0$ limit that the horizon
has the following metric

\begin{equation}
ds^{2}_{h}=
-l^{2} d\xi^{2}_{n} 
-\frac{v^{2}}{4} \left( d\theta_{n}^{2}
+\sin^{2}{\theta_{n}} d\psi_{n}^{2} \right) \, .
\end{equation}

\noindent
with the topology $S^{1} \times S^{2}$, so the solution is a black ring with
non-Abelian hair, \textit{i.e.}~a non-Abelian black ring. Using this metric we
can compute the area of the horizon\footnote{Notice that $\xi_{n} \in
  [0,2\pi)$, as can be deduced from expression (\ref{eq:coordhorizon})
  together with $\int d\Omega_{(3)}=2\pi^{2}$.},

\begin{equation}
\frac{A_{h}}{2\pi^{2}} = \frac{1}{2\pi^{2}}\int d^{3} x \sqrt{\vert g_{h} \vert}=l v^{2}  \, ,
\end{equation}

\noindent
so the entropy of the non-Abelian black ring can be written in terms of the charges and angular momenta using the expressions for the
constants $v$ and $l$ Eqs.~(\ref{eq:v}) and (\ref{eq:l}) together with Eq.~(\ref{eq:modu}) as follows:

\begin{eqnarray}
S 
& = &
\pi \left[ 3 \cdot 6^{2} C^{ijk} C_{klm} 
\left( Q_{i}-C_{ihn} q^{h} q^{n} \right) 
\left( Q_{j}-C_{jpq} q^{p} q^{q} \right) q^{l} q^{m}
- 3\left( q^i Q_i-C_{ijk} q^i q^j q^k \right)^2
  \right. 
\nonumber \\
& & \nonumber \\
& &
\left. \hspace{1cm} 
- \frac{4}{\sqrt{3}}(J_{\phi_{2}} -J_{\phi_{1}}) 
\left( C_{ijk}q^{i}q^{j}q^{k}-16\frac{q^{0}}{g^{2}} \right)
-  3\left( Q_{1}- \frac{q^{0} q^{1}}{3}  \right)^{2} 
\left( \frac{32}{g^{2}} \right) \right]^{1/2} \, .
\end{eqnarray}

%
%
%

Finally, we would like to compute the mass of the solution. We do so by
comparing the asymptotic behavior of the metric component $g_{tt}$ with that
of the Schwarzschild solution, $g_{tt} \sim
1-\frac{8MG}{3\pi\rho^{2}}+\cdots$. We get

\begin{equation}
\label{eq:mass1}
M = 
\frac{3^{5/2} \lambda_{1}}{2} 
\left( \lambda_{1} Q_{0}+2\lambda_{0} Q_{1} \right) \, .
\end{equation}

The constants $\lambda_{i}$ can be expressed in terms of physical constants.
If we define the physical scalars of the theory as $\phi^{x}\equiv
h_{x}/h_{0}$ we find that the only scalar with a non-vanishing asymptotic
value is the Abelian one and this value is
$\phi^{1}_{\infty}=\lambda_{1}/\lambda_{0}$. On the other hand, the asymptotic
normalization of the metric Eq.~(\ref{eq:normalization}) implied $\lambda_{0}
\lambda_{1}^{2}=2/3^{3}$. Then,

\begin{equation}
\lambda_{0}
=
2^{1/3} 3^{-1} \left( \phi^{1}_{\infty} \right)^{-2/3} \, , 
\hspace{1cm} 
\lambda_{1}
=
2^{1/3} 3^{-1} \left( \phi^{1}_{\infty} \right)^{1/3} \, .
\end{equation}

\noindent
and $M$ takes the form

\begin{equation}
\label{eq:mass2}
M = 
2^{-1/3} 3^{1/2} \left[ \left( \phi^{1}_{\infty} \right)^{2/3} Q_{0} 
+ 2 \left( \phi^{1}_{\infty} \right)^{-1/3} Q_{1} \right] \, ,
\end{equation} 

\noindent
and depends only on the moduli and on the electric charges $Q_{0},Q_{1}$ while
the $q^{i}$, which correspond to magnetic dipole momenta do not contribute to
it \cite{Emparan:2004wy}. The non-Abelian field do not contribute, either.

This expression looks identical to that of the non-Abelian black hole solution
constructed in Ref.~\cite{Meessen:2015enl}, but the charges $Q_{0}$ and
$Q_{1}$ are not the same than the charges $q_{0}$ and $q_{1}$ that appear in
the black-hole mass formula given in that reference. They are, actually,
related by $Q^{\rm BR}_{i}=q^{\rm BH}_{i} +C_{ijk}q_{\rm BR}^{j} q_{\rm
  BR}^{k}$. This is just reflecting the fact that the conserved electrical charges in the black ring receive contributions from the magnetic dipole momenta via the Chern-Simons term in the action. This effect is commonly described as "charges dissolved in fluxes" \cite{Bena:2004de}.

This non-Abelian black-ring mass formula, is, however, identical to
that of the Abelian black ring that one would obtain by removing the
non-Abelian fields from this solution. In other words: the presence of
non-Abelian fields is not observable at spatial infinity. They do contribute
to the entropy, though, as in the black-hole case, their entropy being smaller
than that of their Abelian siblings.

\section{Non-Abelian Rotating Black Holes}
\label{sec-rotatingbhs}

In the $R\rightarrow 0$ limit, several things happen:

\begin{enumerate}
\item All the harmonic functions are now centered at $r=0$ (except for $M$
  which becomes constant):

\begin{equation}
H=N=\frac{1}{r} \, , 
\hspace{.5cm}
M=\tfrac{3}{4}\lambda_{i}q^{i}\, ,
\hspace{.5cm}
\Phi^{i}=-\frac{q^{i}}{4\sqrt{2}} N \, ,
\hspace{.5cm}
L_{i}=\lambda_{i}+\frac{Q_{i}-C_{ijk}q^{j}q^{k}}{4} H \, .
\end{equation}

\item The non-Abelian gauge field is also centered at $r=0$:

\begin{equation}
\Phi^{\alpha}
=
\frac{1}{gr\left( 1+\lambda^{2} r \right)} \delta^{\alpha}_{s+1} \frac{x^{s}}{r} \, ,
\hspace{1cm}
\breve{A}^{\alpha} 
= 
\frac{1}{g r \left(1+\lambda^{2} r \right)} \epsilon^{\alpha}{}_{rs} 
\frac{x^{s}}{r} d x^{r} \, ,
\end{equation}

\noindent
and the distorted BPST instanton is not distorted anymore.

\item The metric function $\hat{f}$ is now given by 

\begin{equation}
\hat{f}^{-3}
=
\left[ 
\tfrac{3}{2}\left(\lambda_{0} +\frac{Q_{0}}{4r}\right)
-\frac{2}{g^{2}}\frac{1}{r(1+\lambda^{2}r)^{2}}
\right]
\left[ 
9\left(\lambda_{1}+\frac{Q_{1}}{4r} \right)^{2} 
-\frac{2(q^{0})^{2}}{g^{2}}\frac{1}{r^{2}(1+\lambda^{2}r)^{2}}
\right]\, .    
\end{equation}

\noindent
The mass of this object is identical to that of the black ring
Eqs.~(\ref{eq:mass1}) and (\ref{eq:mass2}). it has no non-Abelian
contributions. The near-horizon limit, though, includes non-Abelian terms

\begin{equation}
\label{eq:fnear0}
\hat{f}^{-1} \sim \frac{Y}{r}\, ,
\,\,\,\,
\mbox{with}
\,\,\,\,
Y^{3} = 
\left(\tfrac{3}{8}Q_{0} -\frac{2}{g^{2}} \right)  
\left(\tfrac{9}{16}Q_{1}^{2}-\frac{2}{g^{2}}(q^{0})^{2} \right)\,   
\end{equation}

\item $\omega$ vanishes identically and $\hat{\omega}$ is determined only by
  $\omega_{5}$, which takes the form

\begin{equation}
\begin{array}{rcl}
\hat{\omega}
& = &
\omega_{5}(d\varphi +\cos{\theta}d\psi)\, ,
\\
& & \\
\omega_{5}
& = &
{\displaystyle
\frac{q^{i}}{16}
\left(3Q_{i}-C_{ijk}q^{j}q^{k}\right)\frac{1}{r}
-\frac{2q^{0}}{g^{2}} 
\frac{1}{r \left(1+\lambda^{2}r\right)^{2}}
}
\, .
\\
\end{array}
\end{equation}

\noindent
As a result, the two angular momenta become identical

\begin{equation}
J_{\phi_{1}} =  J_{\phi_{2}} = \tfrac{1}{2\sqrt{3}}q^{i}
\left(3Q_{i}-C_{ijk}q^{j}q^{k} \right) \equiv J\, .  
\end{equation}

Observe that the non-Abelian term in $\omega_{5}$, which does not contribute
to the angular momentum, does contribute to the $r\rightarrow 0$ limit just as
the Abelian terms:

\begin{equation}
\label{eq:05near0}
\omega_{5}\sim Z/r\, ,
\,\,\,\,  
\mbox{where}
\,\,\,\,
Z 
=
\tfrac{\sqrt{3}}{8}J -\frac{2q^{0}}{g^{2}}\, . 
\end{equation}

\end{enumerate}

Let us study the near-horizon limit $\rightarrow 0$. Using
Eqs.~(\ref{eq:fnear0}) and (\ref{eq:05near0}), we find that the metric
Eq.~(\ref{eq:themetric}) behaves in this limit as

\begin{equation}
ds^{2}
\sim 
\frac{r^{2}}{Y^{2}}dt^{2} -\frac{Y}{r^{2}}dr^{2} -Yd\Omega^{2}_{(2)}
+\frac{2Z}{Y^{2}}rdt (d\varphi +\cos{\theta}d\psi)
+\left(\frac{Z^{2}}{Y^{2}}-Y\right)(d\varphi +\cos{\theta}d\psi)^{2}\, ,
\end{equation}

\noindent
which can be rewritten in the form 

\begin{equation}
ds^{2} 
\sim 
Yd\Pi^{2}_{(2)} - Yd\Omega^{2}_{(2)}  
-Y\left[\sin{\alpha}\rho dt 
-\cos{\alpha}(d\varphi +\cos{\theta}d\psi) \right]^{2}\, ,
\end{equation}

\noindent
where $r=(Y^{3}-Z^{2})^{1/2}\rho$,
$d\Pi^{2}_{(2)}=\rho^{2}dt^{2}-\frac{d\rho^{2}}{\rho^{2}}$ is the metric of
the AdS$_{2}$ of unit radius and $\sin^{2}{\alpha}=Z^{2}/Y^{3}$. This space is
the near-horizon limit of the BMPV black hole \cite{Breckenridge:1996is}, but,
due to the non-Abelian contribution to $Z$ (which can be understood as a sort
of ``near-horizon angular momentum''), now $\alpha$ does not vanish for
vanishing asymptotic angular momentum $J$ and we can have a stationary black
hole with $J=0$ whose near-horizon limit is not AdS$_{2}\times$S$^{3}$. The
converse is also possible: we can make $\alpha=Z=0$ for
$J=\frac{16}{\sqrt{3}}q^{0}/g^{2}$ and have a rotating black hole whose
near-horizon limit is AdS$_{2}\times$S$^{3}$.

The area of the horizon is 

\begin{equation}
\frac{A}{2\pi^{2}} = 8\sqrt{Y^{3}-Z^{2}}\, .  
\end{equation}

\section{Conclusions}
\label{sec-conclusions}

The existence of black-hole and black-ring solutions with identical asymptotic
behaviour but with non-Abelian hair that contributes to the entropy
\cite{Meessen:2008kb,Bueno:2014mea,Meessen:2015nla,Meessen:2015enl} challenges
our understanding of black-hole hair and the microscopic interpretation of the
black-hole/black-ring entropy, just as the Abelian hair discovered in
Ref.~\cite{Emparan:2004wy} did. More research is necessary to gain a better
understanding of these solutions. In particular, the stability of these
supersymmetric non-Abelian solutions (which are entropically disfavored) needs
to be addressed and their possible non-supersymmetric and non-extremal
generalizations have to be constructed and studied. Work in these directions
is in progress.

\section*{Acknowledgments}

PFR would like to thank C.S.~Shahbazi for interesting conversations and the
IPhT for its hospitality. This work has been supported
in part by the Spanish Ministry of Science and Education grant
FPA2012-35043-C02-01, the Centro de Excelencia Severo Ochoa Program grant
SEV-2012-0249 and the Spanish Consolider-Ingenio 2010 program CPAN
CSD2007-00042.  The work of PFR was further supported by the \textit{Severo
  Ochoa} pre-doctoral grant SVP-2013-067903, the EEBB-I-16-11563 grant and the
John Templeton Foundation grant 48222.  TO wishes to thank M.M.~Fern\'andez
for her permanent support.


\end{document}